\documentclass[%
 reprint,
groupedaddress,
amsmath,
amssymb,
aps,
prd,
twocolumn,preprintnumbers,floatfix,nofootinbib]{revtex4-2}

\usepackage[pdftex]{graphicx}
\usepackage{dcolumn}
\usepackage{bm}
\usepackage{hyperref}

\newcommand{\be}{\begin{equation}}
\newcommand{\ee}{\end{equation}}
\newcommand{\bea}{\begin{eqnarray}}
\newcommand{\eea}{\end{eqnarray}}

\usepackage{mathrsfs}
\usepackage{amsmath, amsthm, amssymb}
\usepackage{color}
\usepackage{epstopdf}
\usepackage{amssymb}
\usepackage{verbatim}
\usepackage{enumerate}
\usepackage{float}
\usepackage[caption = false]{subfig}
\usepackage[normalem]{ulem}  
\usepackage{epsfig}
\usepackage{cleveref}

\newcommand{\apjl}{ApJLett}		
\newcommand{\mnras}{MNRAS}		
\newcommand{\jcap}{JCAP}		

\newcommand{\physrep}{Phys.~Rep.}   

\begin{document}
\title{Relaxation times for Bose-Einstein condensation by self-interaction and gravity}

\author{Jiajun Chen$^{1,2}$}
\thanks{They contribute equally to this work. }
\email{chenjiajun@swu.edu.cn}
\author{Xiaolong Du$^3$}
\thanks{They contribute equally to this work. }
\email{xdu@carnegiescience.edu}
\author{Erik W. Lentz$^{2,4}$}
\email{erik.lentz@pnnl.gov}
\author{David J. E. Marsh$^5$}
\email{david.j.marsh@kcl.ac.uk}

\affiliation{
$^1$School of Physical Science and Technology, Southwest University, Chongqing 400715, China\\
$^2$Institut f\"ur Astrophysik, Georg-August-Universit\"at G\"ottingen, Friedrich-Hund-Platz 1, D-37077 G\"ottingen, Germany\\
$^3$Carnegie Observatories, 813 Santa Barbara Street, Pasadena, CA 91101, U.S.A\\
$^4$Pacific Northwest National Laboratory, Richland, WA 99354, U.S.A.\\
$^5$Theoretical Particle Physics and Cosmology, King’s College London, Strand, London WC2R 2LS, United Kingdom}

\date{\today}

\begin{abstract}
In this paper, we study the Bose-Einstein condensation of a scalar field with an attractive self-interaction, with or without gravitational interactions. We confirm through full dynamical simulation that the condensation timescale due to self-interaction is inversely proportional to  the square of the number density $n$ and the self-coupling constant $g$ : $\tau \propto n^{-2} g^{-2}$. We also investigate the condensation timescale when self-interaction and gravity are both important by solving the Gross-Pitaevskii-Poisson equations, and find that the condensation time scales according to an additive model for the cross section. We discuss the relevance of our results to theoretical models of boson star formation by condensation.
\end{abstract}

\maketitle

\section{Introduction}\label{sec:intro}

The composition of dark matter (DM) is one of the most important unresolved problems in modern cosmology.
Observations show DM makes up about $27\%$ of the total mass–energy in our Universe \cite{Aghanim:2018eyx}, to which many different models for dark matter particles have been posited.
One promising idea is that dark matter is composed of light bosons \cite{pecceiquinn1977,weinberg1978,wilczek1978,Kim:1979if,Shifman:1979if,Dine:1982ah,Zhitnitsky:1980tq,1981PhLB..104..199D, Arkani_Hamed_2009,2015JCAP...02..006P,Marsh:2015xka,PhysRevLett.123.021102,PhysRevD.101.096013,Chen:2020cef}, such as the ultra-light axion-like particles with masses $10^{-22}-10^{-19}$ eV \cite{1990PhRvL..64.1084P,2000PhRvD..62j3517S,hu2000,2000ApJ...534L.127P,amendola2005,grin2019gravitational} or QCD axions with masses $10^{-11}-10^{-2}$ eV \cite{Marsh:2015xka,Chen:2020cef,Kolb:1994fi}. It has been shown that these light bosons can condense and form compact objects balanced by competing dynamics of self-interactions, gravity and gradient energy. We call them solitons or boson stars\cite{1968PhRv..172.1331K,PhysRevD.42.384,1993PhRvL..71.3051K,Widrow:1993qq,2015MNRAS.451.2479M, Hui:2016ltb}.

Observational signatures and phenomenology of boson stars is an active topic \cite{2011PhRvD..84d3531C,Chavanis:2011zm,Maga_a_2012,Schive:2014hza,2015MNRAS.451.2479M,Levkov:2016rkk}. In 2018, \citet{Levkov:2018kau} studied the condensation of boson stars by gravitational interaction from an isotropic initial distribution of particles. They also gave the theoretical prediction for the relaxation time of bosons by self-interaction, $\tau_{\rm self} \propto 1/(n^2 |g|^2)$, and compared it with the relaxation time by gravitational interaction, but did not directly simulate the case with self-interactions~\citep{Levkov:2018kau}. In our previous work, we used pseudospectral methods with full three-dimensional non-linearity~\cite{Mocz:2017wlg,Du:2018qor,Levkov:2018kau,Chen:2020cef} to study the condensation and growth of boson stars both with and without self-interactions~\cite{Chen:2020cef}. The present work builds on our previous simulations and directly measures the effect of self-interactions on the condensation time.

We find the following results in our simulations:
\begin{itemize}
\item The condensation time of boson stars with attractive self-interaction is inversely proportional to $n^2|g|^2$ as predicted by \citet{Levkov:2018kau}.
\item The multi-physics condensation time of boson stars fits the analytic formula $\tau_{\rm gravity}\tau_{\rm self}/(\tau_{\rm gravity}+\tau_{\rm self})$ for bosons with gravity and attractive self-interaction.
\end{itemize}

The remainder of this paper is structured as follows. Section~\ref{GPP_Equations} introduces the Gross-Pitaevskii-Poisson (GPP) equations and initial conditions used for simulation. Section~\ref{sec:theory} presents theoretical predictions for the condensation time, while Section~\ref{sec:numerics} studies the relaxation time through numerical simulations. Lastly, a discussion of our conclusions is presented in Section~\ref{Conclusion}.

\section{The Gross-Pitaevskii-Poisson equations and Initial conditions}
\label{GPP_Equations}
When the occupation number of a light real scalar field is very large, it can be described by a classical field $\phi$\cite{Tkachev:1991ka,PhysRevD.61.083517}.
The self-interaction potential of the scalar field \cite{Sikivie:2006ni,Arvanitaki:2009fg,2017PhRvL.118a1301L} can be expanded for small field values relative to a scale factor $f_a$ as 
\begin{equation}
V(\phi) = \frac{1}{2} m^2 \phi^2 \pm \frac{1}{4!}\frac{m^2}{f_a^2} \phi^4+...,
\end{equation}
where $m$ and $f_a$ are the particle mass and the decay constant, respectively. Here we use natural units: $\hbar=c=1$. Furthermore, we define the dimensional self-coupling constant as $g \equiv \pm \frac{1}{8 f_a^2}$. The value of the particle mass, $m$, and coupling constant, $g$, depend on the detailed models. 

In the non-relativistic, low-velocity and low-density limits, we can write $\phi$ as
\begin{equation}
\phi= \sqrt{\frac{2}{m}} {\rm Re}\left[\psi e^{-i m t}\right].
\label{eq:phi_psi}
\end{equation}
At lowest order in field intensity, the complex wave function $\psi$ satisfies the Gross-Pitaevskii-Poisson (GPP) equations
\cite{Chavanis:2011zm,Eby:2015hsq}
\begin{eqnarray}
i\frac{\partial}{\partial{t}}\psi&=&-\frac{1}{2m}\nabla^2\psi + m V\psi+g|\psi|^2\psi,
\label{eq:GPP1}
\\
\nabla^2{V}&=&4 \pi G m\left(|\psi|^2-n\right),
\label{eq:GPP2}
\end{eqnarray}
where $n$ is the mean number density,  $G$ is Newton's gravitational constant, and $V$ is the gravitational potential.
Introducing the dimensionless quantities
\begin{eqnarray}
x=\widetilde{x}/(m v_0),\quad t=\widetilde{t}/(mv_0^2),\quad V=\widetilde{V} v_0^2,\nonumber\\
\psi=\widetilde{\psi}v_0^2\sqrt{m/(4 \pi G)},\quad g=\widetilde{g}\,4\pi G/v_0^2,
\label{eq:dim}
\end{eqnarray}
where $v_0$ is a reference velocity (e.g. the characteristic velocity of the initial state), we obtain the dimensionless equations
\begin{eqnarray}
i\frac{\partial}{\partial{\widetilde{t}}}\widetilde{\psi}&=&-\frac{1}{2}\widetilde{\nabla}^2\widetilde{\psi} + \widetilde{V}\widetilde{\psi}+\widetilde{g}|\widetilde{\psi}|^2\widetilde{\psi},
\label{eq:GPP1_dim}
\\
\widetilde{\nabla}^2{\widetilde{V}}&=&|\widetilde{\psi}|^2-\widetilde{n}.
\label{eq:GPP2_dim}
\end{eqnarray}

For initial conditions, we test both the Dirac delta $|\psi_{\vec{p}}|^2=N\delta(|\vec{p}|-mv_0)$ and the Gaussian  $|\psi_{\vec{p}}|^2= Ne^{-p^2}$ momentum distributions \cite{Levkov:2018kau}.
The simulation is performed in a periodic box of size $\widetilde{L}$, the total number of non-relativistic bosons in the box is $N\equiv n L^3$. Performing an inverse Fourier transform on $|\psi_{\vec{p}}| e^{i S}$ with random phases, $S$, we obtain an isotropic and homogeneous initial distribution in position space, $\psi(\vec{x},0)$. The box size is chosen to be $\widetilde{L}<2\pi/\widetilde{k}_J$, where $\widetilde{k}_J=(4\widetilde{n})^{1/4}$ is the dimensionless Jeans wavenumber, so that there is not also a halo formed in the box.
To study the influence of self-interaction, we vary the dimensionless coupling constant $\widetilde{g}$ over the range $[-100,0]$.

\section{Theoretical Prediction for the Condensation Time of Boson Stars}\label{sec:theory}

The condensation time of boson
stars, $\tau$, is expected to be proportional to the kinetic relaxation time, which takes the form
\begin{equation}
    \tau =\frac{2\sqrt{2} b m^3 v^2}{3\sigma n^2 \pi^2},
\end{equation}
where $v$ is the characteristic velocity, $n$ is the average density of systems, $\sigma$ is the transport cross section of interaction, and $b$ is a constant depending on the initial configuration.

For the gravitational interaction, the  cross section is $\sigma_{\rm gravity} =8\pi(mG)^2\log(mvR)/v^4$. Thus, the condensation time of boson stars by gravity is 
\begin{equation}
    \tau_{\rm gravity} =\frac{2\sqrt{2} d m^3v^2}{3\sigma_{\rm gravity} n^2 \pi^2}  = \frac{b\sqrt{2}}{12\pi^3}\frac{mv^6}{G^2n^2\log(mvR)}.
    \label{eq:condensation_time}
\end{equation}
This prediction has been tested by a large number of numerical simulations~\cite{Eggemeier:2019jsu,Levkov:2018kau,Veltmaat:2019hou,Chen:2020cef}.
Levkov et al. find the coefficient $b \approx 0.7 $ for delta-distributed bosons and $b\approx 0.9$  for Gaussian-distributed bosons~\cite{Levkov:2018kau}.

Levkov et al. also give the relaxation time due to self-interaction, having cross section $\sigma_{\rm self} = m^2 g^2/(2 \pi)$~\cite{Levkov:2018kau}.
The condensation time of boson stars by self-interaction is thus given by:
\begin{equation}
    \tau_{\rm self} = \frac{2\sqrt{2} d m^3v^2}{3\sigma_{\rm self} n^2 \pi^2} = \frac{4\sqrt{2} d m v^2}{3 n^2 g^2 \pi},
    \label{eq:condensation_time_self_g2}
\end{equation}
where $d$ is an $\mathcal{O}(1)$ coefficient to be determined by simulation. What has not been tested to date is how exactly the condensation time of self-interactions depends on the strength of self-coupling, i.e. verifying Eq.~\eqref{eq:condensation_time_self_g2}, and measuring the constant $d$ in simulation. 

Next consider the case with both gravitational and self-interactions. For bosons with both self-interaction and gravity, one might estimate the total condensation rate as being additive in the individual rates: $\sigma_{\rm total} = \sigma_{\rm gravity} + \sigma_{\rm self}$. If we invert the rate to find the total time-scale, defining $\tau_{\rm total} \propto 1/\sigma_{\rm total}, \tau_{\rm gravity} \propto 1/\sigma_{\rm gravity}, \tau_{\rm self} \propto 1/\sigma_{\rm self}$, and rearrange, we obtain
\begin{equation}
    \tau_{\rm total} = \frac{\tau_{\rm gravity}\tau_{\rm self}}{ \tau_{\rm gravity} + \tau_{\rm self}}.
    \label{eq:condensation_time_self_gravity_native}
\end{equation}
We will also investigate this model in by simulation of the full GPP equations with gravity and interactions.


\section{Numerical Results for the Condensation Time of Boson Stars}\label{sec:numerics}
\subsection{Condensation of boson stars with attractive self-interaction}
\label{simulation_self_interaction}
For bosons with weak attractive self-interaction, such as QCD axions and fuzzy dark matter, it is not very easy to know the exact relationship between condensation time and self-interaction due to computational limitations and the dominance of the condensation due to gravity~\cite{Levkov:2016rkk,Kirkpatrick:2020fwd,Chen:2020cef}. Thus, here we first consider bosons with attractive self-interaction much stronger than those expected for axions. Gravity is omitted in this section.

For bosons with only self-interaction ($\widetilde{V} = 0$), the GPP equations can be simplified to the Gross-Pitaevskii (GP) equation:
\begin{eqnarray}
i\frac{\partial}{\partial{\widetilde{t}}}\widetilde{\psi}&=&-\frac{1}{2}\widetilde{\nabla}^2\widetilde{\psi} +\widetilde{g}|\widetilde{\psi}|^2\widetilde{\psi}.
\label{eq:GP}
\end{eqnarray}
In order to test 
whether
Eqs.~\eqref{eq:condensation_time_self_g2}
is correct and understand the relaxation time of bosons by self-interaction exactly, numerous simulations are performed of bosons with attractive self-interaction to measure the condensation time to form a boson star.

\begin{figure}[htbp]
\centering
\includegraphics[width=\columnwidth]{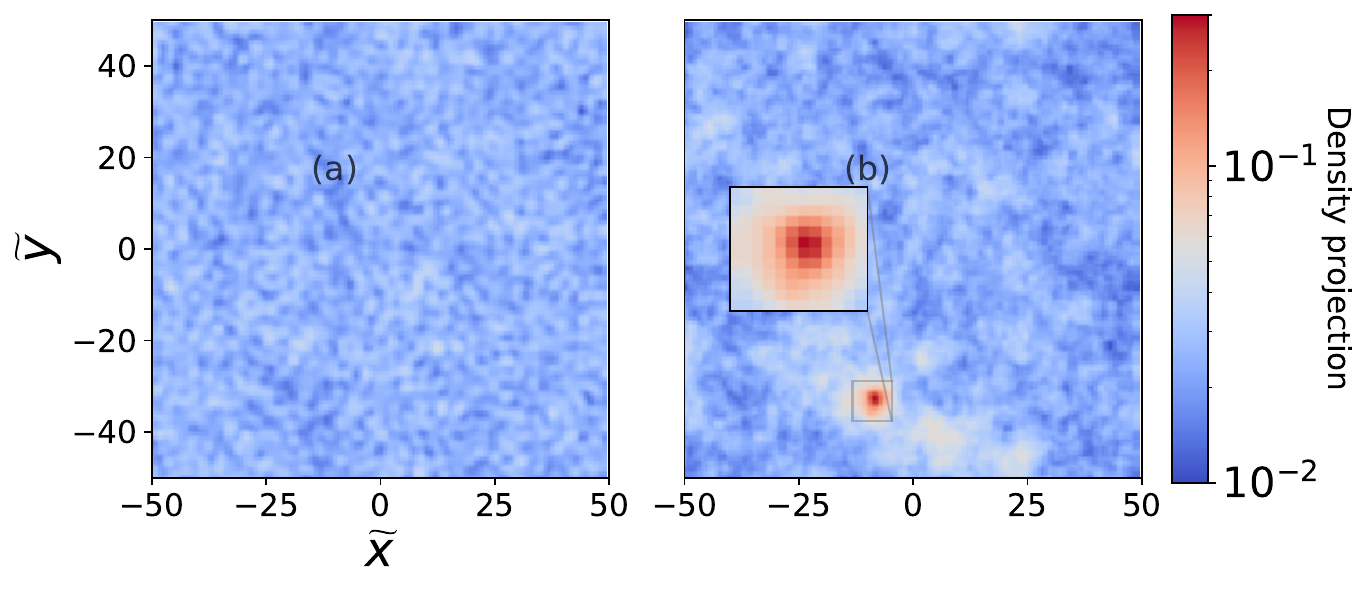}
\caption{Snapshots of the density field from one simulation with box size $\widetilde{L}=100$, total mass $\widetilde{N}=251.2$ and an attractive self-interaction coupling $\widetilde{g} = -13.0$ . (a) Projected density at the initial time. (b) Projected density at $\widetilde{t}=57800$, which shows that a boson star is forming in the box.}
\label{fig:density_projection_self}
\end{figure}

\begin{figure}[htbp]
\centering
\includegraphics[width=\columnwidth]{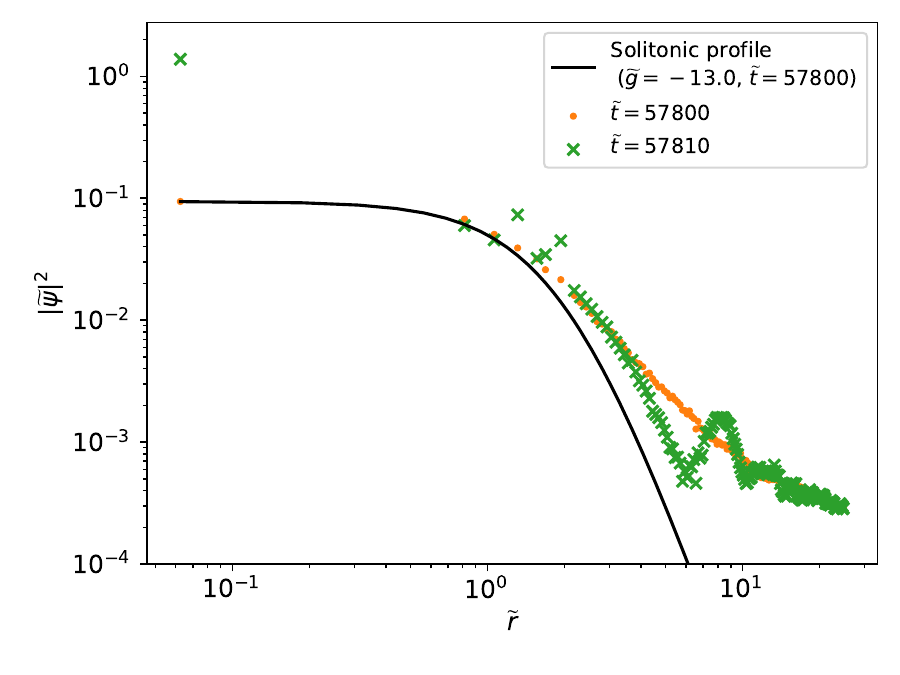}
\caption{Density profiles of the dense core from simulations (colored dots), compared with solitonic profile (solid line) with attractive self-interaction with the same central density \cite{Chen:2020cef}.}
\label{fig:densityprofile_self}
\end{figure}

\begin{figure}[htbp]
\centering
\includegraphics[width=\columnwidth]{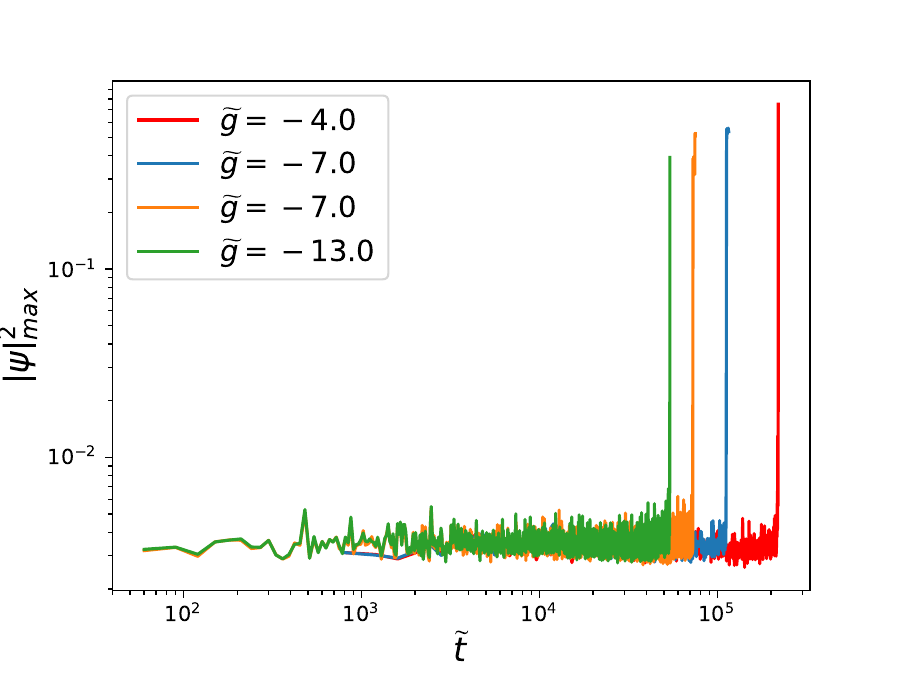}
\caption{The maximum density evolution of bosons with attractive self-interaction for box size $\widetilde{L}=100$, total mass $\widetilde{N}=251.2$ and self-interaction coupling $\widetilde{g} = -13.0$, $\widetilde{g} = -10.0$, $\widetilde{g} = -7.0$, and $\widetilde{g} = -4.0$. The spike at the end of each run indicates the formation of a boson star.}
\label{fig:maxdensity_self}
\end{figure}

Numerically solving the GP equation with different $\widetilde{L}$ and $\widetilde{N}$, we observe the formation of boson stars. One example is shown in Fig.~\ref{fig:density_projection_self} for the box size $\widetilde{L}=100$ and total mass $\widetilde{N}=251.2$. We can see a dense object forms at $\widetilde{t} \approx 57800$.  The density profile of the object is shown in Fig. \ref{fig:densityprofile_self}. The profile from simulation (colored dots) does not deviates from the solitonic profile (solid line) since the boson star grows so fast that it does not reach a quasi-equilibrium state. After that time, the maximum density of boson star grows rapidly, which means it collapses \cite{2011PhRvD..84d3531C,Chavanis:2011zm,Chavanis:2016dab,Levkov:2016rkk,Chen:2020cef}, see Figs.~\ref{fig:densityprofile_self} and \ref{fig:maxdensity_self}. 
We also find that the condensation time, $\tau_{\rm self}$, decreases with the increase of absolute value of coupling constant squared, $|g|^2$.

\begin{figure}[htbp]
\centering
\includegraphics[width=\columnwidth]{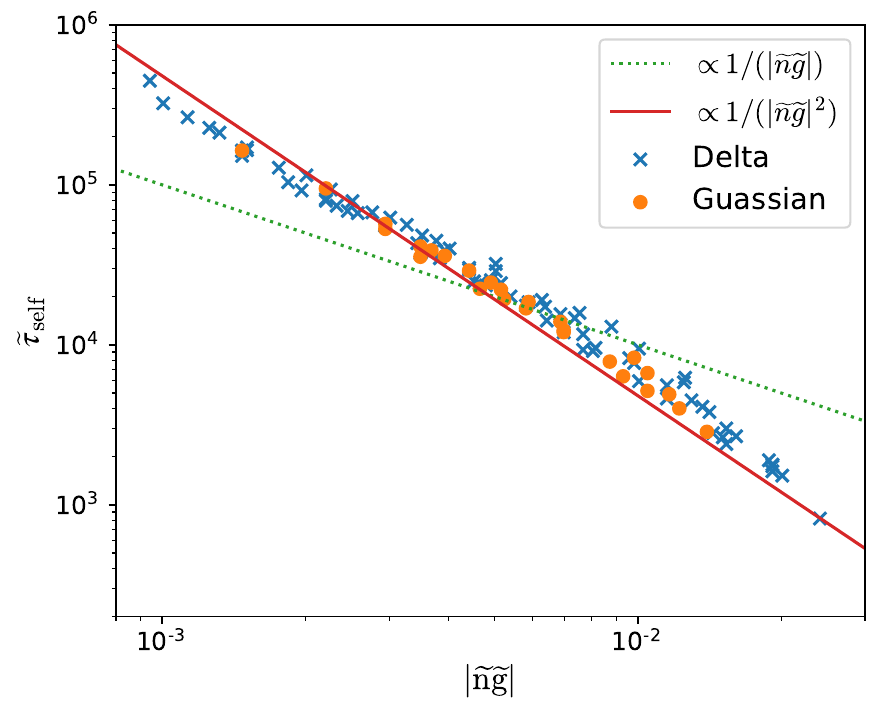}
\caption{Condensation time with respect to $\widetilde{n}|\widetilde{g}|$. Results from both Dirac delta (cross) and Gaussian (filled circles) initial conditions are shown. The simulation results are well-fit by Eq.~\eqref{eq:condensation_time_self_g2}.}
\label{fig:condensation_time_gorg2}
\end{figure}

We repeat such simulations many times with different coupling strengths in order to determine the accurate dependence of the condensation time on $|g|$. We confirm the prediction by Ref.~\cite{Levkov:2018kau}, i.e. Eq.~\eqref{eq:condensation_time_self_g2}, see Fig. \ref{fig:condensation_time_gorg2}. The best-fit value of the coefficient $d$ is $0.8$. For comparison, we show also the scaling $\tau\propto 1/(|\tilde{n}\tilde{g}|)$, which is a poor fit to the data.

\subsection{Condensation of boson stars with both self-interaction and gravity}
\label{simulation_gravity_self}

\begin{figure}[htbp]
\centering
\includegraphics[width=\columnwidth]{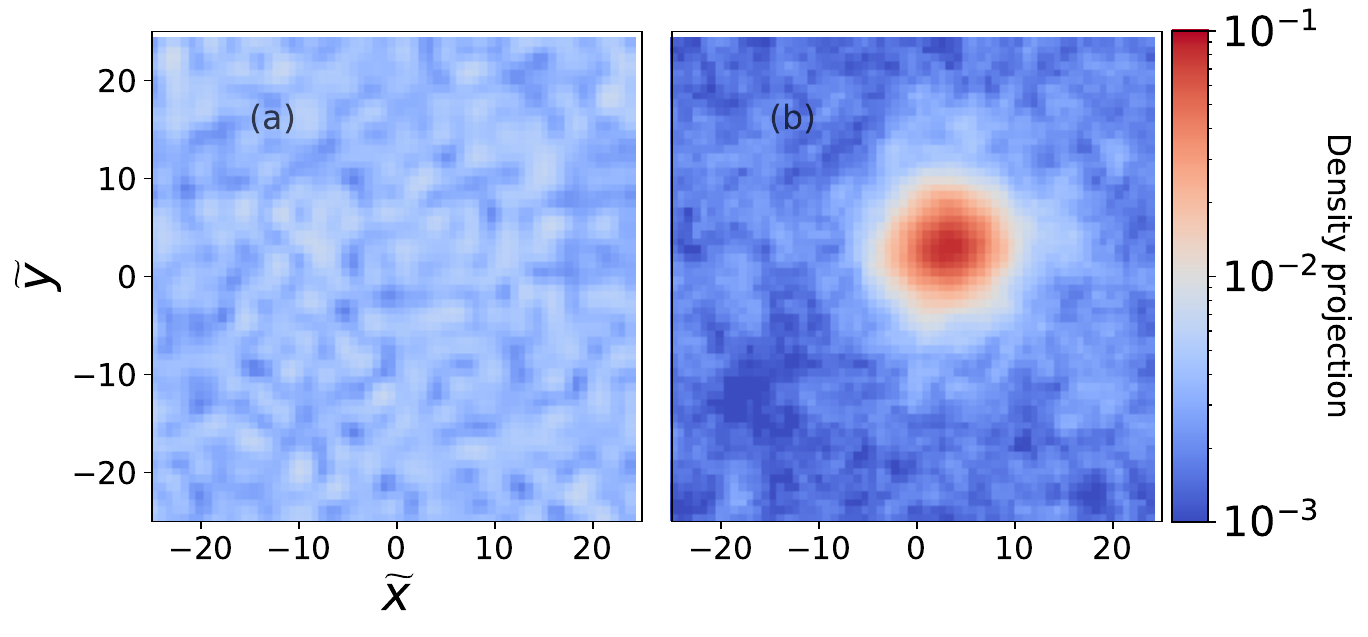}
\caption{Snapshots of the density field from one simulation for bosons with gravity and attractive self-interaction with box size $\widetilde{L}=50$, total mass $\widetilde{N}=10$ and self-interaction coupling $\widetilde{g} = -8.0$. (a) Projected density at the initial time. (b) Projected density at $\widetilde{t}=673500$, which shows that a boson star is forming in the box.}
\label{fig:density_projection_self_gravity}
\end{figure}

\begin{figure}[htbp]
\centering
\includegraphics[width=\columnwidth]{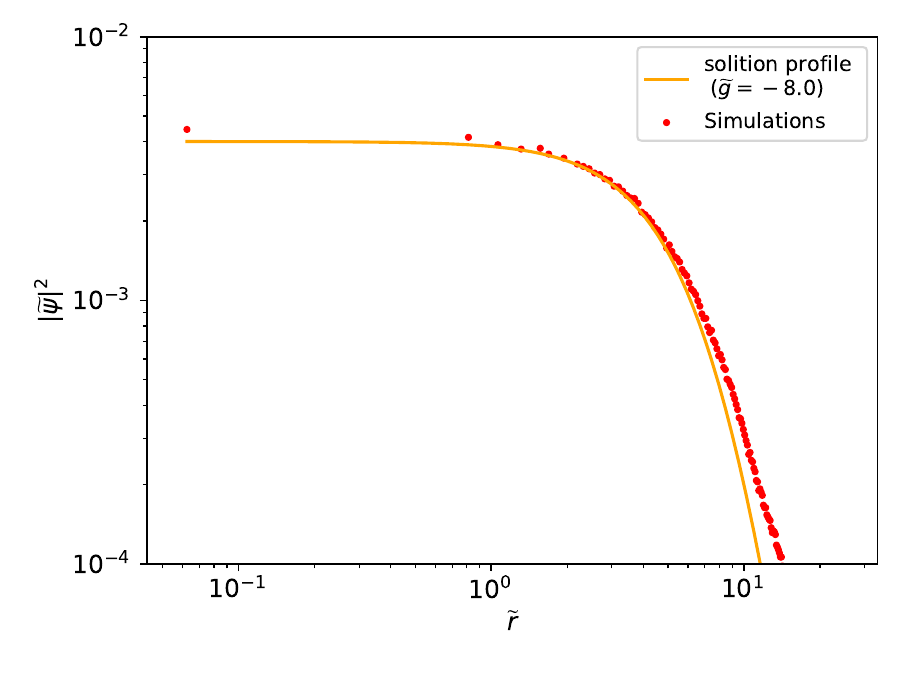}
\caption{Density profiles of the dense core from simulations (colored dots), compared with solitonic profile (solid line) with gravity and attractive self-interaction with the same central density \cite{Chen:2020cef}.}
\label{fig:densityprofile}
\end{figure}

\begin{figure}[htbp]
\centering
\includegraphics[width=\columnwidth]{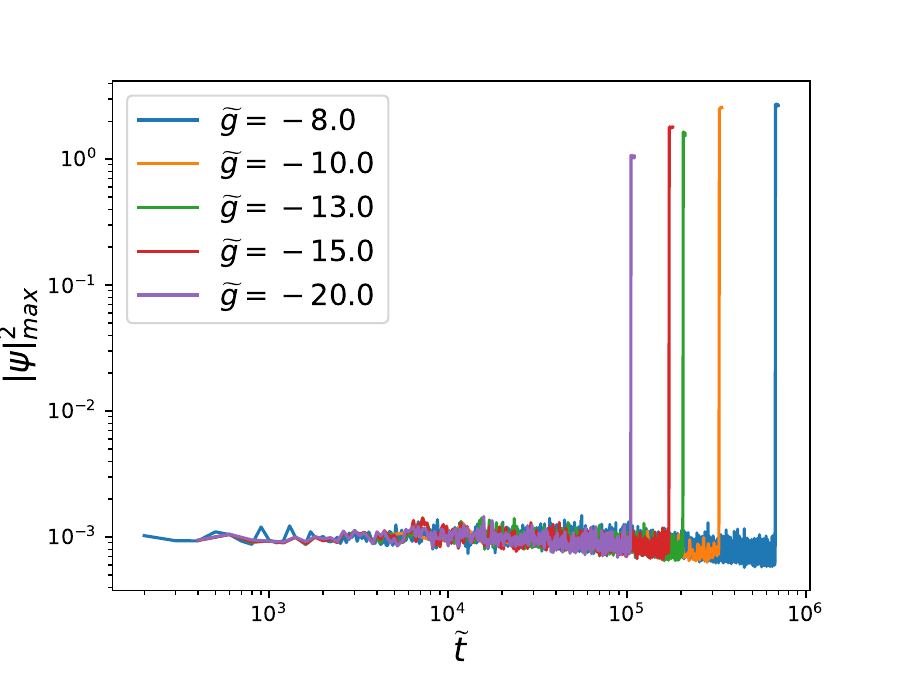}
\caption{The maximum density evolution of bosons with both gravity and attractive self-interaction for total mass $\widetilde{N}=10$, box size $\widetilde{L}=50$ and self-interaction coupling $\widetilde{g} = -20.0$, $\widetilde{g} = -15.0$, $\widetilde{g} = -13.0$, $\widetilde{g} = -10.0$, and $\widetilde{g} =8.0$. The spike at the end of each run indicates the formation of a boson star.}
\label{fig:maxdensity_attself_gravity}
\end{figure}

\begin{figure}[htbp]
\centering
\includegraphics[width=\columnwidth]{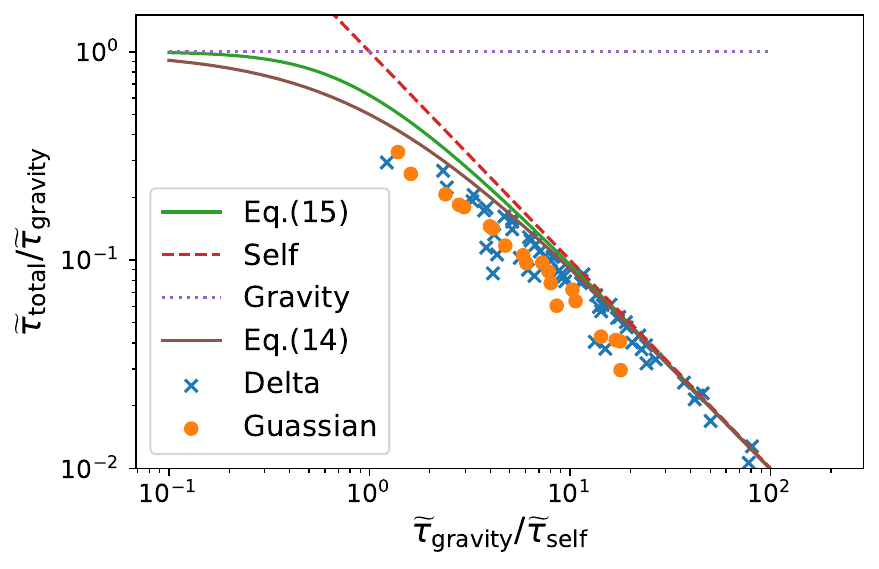}
\caption{Condensation time of bosons with both attractive self-interaction and gravity. Both Dirac delta (red) and Gaussian (blue) initial conditions are shown. The solid lines are  given by Eqs.~\eqref{eq:condensation_time_self_gravity_native} (purple).
}
\label{fig:condensation_time_total}
\end{figure}

We now obtain the condensation time of boson stars by numerically solving the GPP equations, i.e. allowing self-gravity.
The box size for these simulations is in the range $20<\widetilde{L}<150$ and the total mass is in the range $5<\widetilde{N}<2000$. A typical simulation, with the box size $\widetilde{L}=50$ and total mass $\widetilde{N}=10$, is shown in Fig.~\ref{fig:density_projection_self_gravity}. A dense object is formed at $\widetilde{t} \approx 673500$.
The density profile of this object closely follows that of the soliton with gravity and attractive self-interaction\cite{Chen:2020cef}, see Fig.~\ref{fig:densityprofile}. After condensation, the boson star again continues to collapse with increasingly rapid growth, see Fig.~\ref{fig:maxdensity_attself_gravity}.
The condensation time obtained from simulations is compared with Eq.~\eqref{eq:condensation_time_self_gravity_native} as is shown in Fig. \ref{fig:condensation_time_total}. We observe that the simulation results are in agreement with Eq.~\eqref{eq:condensation_time_self_gravity_native} .

\section{Conclusions}
\label{Conclusion}
 
By means of numerical simulation of the dynamical Gross-Pitaevskii-Poisson equations, we studied the condensation time of boson star subject to self-interactions and gravity. For ease of simulation, we studied the cases in which the relaxation timescale for self-interactions is comparable to the timescale for gravity.
In this regime, boson stars can form and collapse to  axion novas or black holes within the lifetime of the Universe due to their self-gravitation and self-interaction, although our non-relativistic simulations cannot study this physics directly. When interaction-driven collapse occurs, it is very fast, making it difficult to render a stable density profile that precisely fits the theoretical profile. However, the formation of a solition is clear.

In the case of bosons with attractive self-interaction, we demonstrated the theoretical prediction of Levkov et al.~\cite{Levkov:2018kau}. We ran $\mathcal{O}(100)$ simulations to confirm the relationship between self-interaction and condensation time. We studied the goodness of fit using the square residuals:
\begin{equation}
\mathcal{SSE} = \sum \left(\frac{\widetilde{\tau}_{\rm self}-\widetilde{\tau}_{\rm self,p}}{\tau_{\rm self}}\right)^2,
\label{eq:SSE}
\end{equation}
where $\tau_{\rm self}$ is measured from simulations and $\widetilde{\tau}_{\rm self,p}$ is the model prediction. The case of $\tau_{\rm self}\propto \frac{1}{|g|^2}$ has $\mathcal{SSE} = 7.07$ and is a much better fit to our results than case of $\tau_{\rm self}\propto \frac{1}{|g|}$ which gives $\mathcal{SSE} = 40.60$. Further, for bosons with gravity and a strong attractive self-interaction, we also find the naive assumption of additive interaction rates provides an excellent fit to our results with $\mathcal{SSE} = 9.70$.

Our result that in the pure self-interaction case $\tau\propto 1/|g|^2$ is in contradiction to the theoretical prediction stated in \citet{Kirkpatrick:2020fwd}, which predicted $\tau\propto 1/|g|$. The theoretical model of Ref.~\cite{Kirkpatrick:2020fwd} substitutes the Schr\"{o}dinger equation potential into the kinetic equation, leading to a result proportional to $1/|g|$. However, the contact interaction mediates no long-range force, and at linear order does not lead to condensation, only time reversal symmetric processes. The correct potential to use in this derivation is the long-range non-relativistic potential, which arises only at one-loop in $\phi^4$ theory, is proportional to $|g|^2$, and is related to the scattering cross section $\sigma_{\rm self}$ by the optical theorem~\cite{Mueller:2002gd} (see also Ref.~\cite{Micha:2004bv} and references therein for the derivation of the $\phi^4$ kinetic equation). Furthermore, Ref.~\cite{Kirkpatrick:2020fwd} gave a prediction for the rate combining gravity and self interactions with a different functional form than the additive rates case, Eq.~\ref{eq:condensation_time_self_gravity_native}. We tested the alternative functional form, and found $\mathcal{SSE} = 15.41$ indicating a worse fit than the additive rates case.

Our results for the accurate condensation time and growth of boson stars can be used 
in boson star population modelling, with applications to the astrophysical phenomenology of bosonic dark matter, e.g. boson star explosions, and merger rates.

\emph{Note Added:} Since the initial release of our manuscript, Kirkpatrick et al. released a second manuscript, Ref.~\cite{Kirkpatrick:2021wwz}, that updates their calculations on the condensation time by self-interaction to account for the upper limit on the momentum occupation in the axion field and arrive at similar conclusions to ours for the scaling of condensation time.

\section{Acknowledgements}
\label{Acknow}
We thank D.G. Levkov for helpful discussions. X. D. acknowledges support from NASA ATP grant 17-ATP17-0120. X.D. thanks A. E. Mirasola for beneficial discussions. E. L. acknowledges funding and support from the U.S. Department of Energy (DOE).  Pacific Northwest National Laboratory is operated by the Battelle Memorial Institute for the U.S. Department of Energy under contract DE-AC05-76RL01830.

\bibliography{Reference}

\end{document}